\documentstyle[twocolumn,seceq,epsf]{jpsj}
\renewcommand{\theequation}{\arabic{equation}}
\catcode`\@=11
\def\simle{\mathrel{\mathpalette\@versim<}}   
\def\simge{\mathrel{\mathpalette\@versim>}}   
\def\@versim#1#2{\lower2.5pt\vbox{\baselineskip0pt \lineskip-.5pt
   \ialign{$\m@th#1\hfil##\hfil$\crcr#2\crcr\sim\crcr}}}
\catcode`\@=12
\def\runtitle{}
\def\runauthor{}

\title
{
Incoherent Charge Dynamics in Perovskite Manganese 
Oxides
}

\author
{ 
Hiroki {\sc Nakano}, Yukitoshi {\sc Motome}$^{1}$ and 
Masatoshi {\sc Imada}
}

\inst
{
Institute for Solid State Physics, University of Tokyo, 
Roppongi 7-22-1, Minato-ku, Tokyo 106-8666\\
$^1$Department of Physics, Tokyo Institute of Technology, Oh-okayama 
2-12-1, Meguro-ku, Tokyo 152-8551
}

\recdate
{
}

\abst
{
A minimal model is proposed for the perovskite manganese oxides 
showing the strongly incoherent charge dynamics 
with a suppressed Drude weight 
in the ferromagnetic and metallic phase near the insulator. 
We investigate a generalized double-exchange model 
including three elements; 
the orbital degeneracy of $e_g$ conduction bands, 
the Coulomb interaction and fluctuating Jahn-Teller distortions. 
We demonstrate that Lancz$\ddot{\rm o}$s diagonalization calculations 
combined with Monte Carlo sampling 
of the largely fluctuating lattice distortions 
result in the optical conductivity which  quantitatively accounts for 
the experimental indications. 
It is found that all the three elements are indispensable 
to understand the charge dynamics in these compounds. 
}

\kword
{
perovskite manganites, double-exchange model, Coulomb interaction, 
orbital degeneracy, Jahn-Teller distortion, metal-insulator transition, 
ferromagnetic metal, optical conductivity, Drude weight, 
incoherent charge dynamics, 
exact diagonalization, strongly correlated electron system
}

\begin{document}
\sloppy
\hyphenation{Hamil-ton-ian}
\maketitle

As an open problem of 
perovskite manganese oxides $R_{1-x}A_{x}$MnO$_3$ 
($R$=rare earth, $A$=Ca, Sr, Ba or Pb)\cite{Ramirez_rev} 
beyond the double-exchange (DE) 
mechanism \cite{Zener,Anderson_Hasegawa,de_Gennes,Kubo_Ohata,Frkw_DMF}; 
strongly incoherent charge dynamics 
in the optical conductivity $\sigma(\omega)$ of a typical compound, 
La$_{1-x}$Sr$_{x}$MnO$_3$, in the ferromagnetic and metallic (FM) phase 
near the insulator ($x$=0.175, 0.18) 
is known\cite{Okimoto_let,Okimoto_reg,Okuda_NEC}. 
The coherent component of $\sigma(\omega)$, 
namely the Drude weight, at low temperatures is suppressed;  
its ratio to the effective carrier density is about 0.2 
and the nominal mass enhancement is very large ($m^{*}/m $=50-80). 
The two-dimensional (2D) system of La$_{1.2}$Sr$_{1.8}$Mn$_2$O$_7$
also reveals the incoherent charge dynamics\cite{Ishikawa}.
Such an incoherent feature cannot originate from 
the spin degrees of freedom under the spin polarization 
in the FM phase.  

Possible elements as candidates contributing to the incoherence 
are orbital degeneracy, electron correlation and 
dynamical Jahn-Teller (JT) distortion, 
which are neglected in the simple DE model.  
A question we are faced with is what is the minimal model 
for the manganites showing such a strong incoherence. 
Previous studies suggest that a partial account of the three elements 
cannot reproduce the dominance of incoherence quantitatively. 
(i) Only the $e_{g}$-orbital degeneracy\cite{Shiba_Shiina_Takahashi} 
leads to an incoherence comparable to the coherent part; 
the ratio of the Drude weight to the total weight is $\sim$0.46. 
(ii) The $e_{g}$-orbital degeneracy combined 
with the JT distortions\cite{Millis,Yunoki} gives 
the ratio $\sim$0.7. 
(iii) For a model containing both the $e_{g}$-orbital degeneracy 
and the electron correlation, a broad incoherent structure
of $\sigma(\omega)$ in the energy region below the charge gap
for the slightly hole-doped case is obtained. 
However, the mean-field treatment\cite{Ishihara_Yamanaka_Nagaosa} gives 
the absent Drude weight. Numerical calculations in 2D 
at substantially high temperature conclude 
the ratio of the Drude weight to the effective carrier density 
to be $\sim$0.4, 
in spite that the coherence is quite suppressed 
due to the thermal fluctuations\cite{Horsch_finite_temp}. 
Here the effective carrier density is defined as the Drude weight 
added by the incoherent weight within the lower Hubbard band. 
Numerical calculations 
at $T=0$\cite{Nakano_Motome_Imada_orbital_U,NEC_Motome_Nakano_Imada}, 
where the quantum fluctuations are fully taken into account, 
give the ratio $\sim$0.45 under the strong Coulomb interaction, 
even when the chemical potential difference between two orbitals 
in a 2D plane is controlled to let the orbital polarization (OP) 
vanish so that the three-dimensional (3D) system is simulated. 
Thus, if only a part of the three elements is taken into account,
the Drude and incoherent weights are at most comparable
and cannot reproduce the dominance of incoherence observed
in charge dynamics of the experiments.

In such circumstances, 
a model containing all the three elements, 
the $e_g$-orbital degeneracy, 
the electron correlation and the JT distortion 
with the staggered and long-ranged pattern of 
the displacement of the oxygens surrounding 
the Mn atoms has been investigated. 
A cooperative effect of these elements are 
known to play an essential role 
for the undoped compounds showing 
an insulating behavior\cite{Motome_Imada_let,Motome_Imada_reg}. 
The results for the doped metal\cite{Nakano_Motome_Imada_LT22} 
suggest that this staggered distortion 
under the strong correlation makes the charge dynamics more incoherent 
because the distortion enhances the orbital fluctuation 
obstructing the motion of electrons.  
In the experiments, on the contrary, 
no long-ranged orders of JT distortion are observed 
at the doping where the strong incoherence appears, 
although short-ranged JT correlations are expected to persist. 
To reach full understanding of the origin of the incoherence, 
it is required to examine 
whether the charge dynamics becomes as incoherent as observed 
in the experiments when we treat such short-ranged 
JT correlations together with the orbital degeneracy 
and the electron correlation.  

When the displacement of the oxygens is considered, 
one should be careful about a zero-point energy (ZPE) 
which comes from the quantum nature of the lattice. 
One can decompose the displacement of the oxygens 
into the normal modes of the lattice, among which 
we consider the $Q_2$ and $Q_3$ modes coupling with 
the electrons in the $e_g$ orbitals through the JT coupling. 
Consequently, one has the phonon part in the Hamiltonian to be 
$\sum_i [-\frac{\hbar^2}{4m_0 a^2} 
(\frac{\partial^2}{\partial \hat{q}_{2i}^2}
+\frac{\partial^2}{\partial \hat{q}_{3i}^2}) 
+ k (\hat{q}_{2i}^2+\hat{q}_{3i}^2)]$ 
where $m_0$ is the mass of an oxygen atom, 
$a$ is the distance between two neighboring Mn atoms, 
$k$ is the elastic constant\cite{comment_coefficient_phonon}. 
$\hat{q}_{2i}$ and $\hat{q}_{3i}$ are quantum variables 
corresponding to the $Q_{2}$ and $Q_{3}$ modes  
at a manganese site $i$, respectively, in the length unit $a$. 
Thus, the ZPE per an Mn site is found to be 
$E_{\rm ZPE}=(\hbar/a)\sqrt{k/m_0}$. 
Since the elastic constant $k$ is roughly estimated from the frequency 
of a phonon of oxygen-bond stretching 
as the order of 10 - 100 eV,
$E_{\rm ZPE}$ is about 0.01 - 0.04 eV, where, 
$m_0$\,$\sim$\,$2.7$\,$\times$\,$10^{-26}$[kg] and 
\mbox{$a$\hspace{0.8pt}$\sim$\hspace{0.8pt}$4$\hspace{0.8pt}${\rm \AA}$} 
in La$_{1-x}$Sr$_{x}$MnO$_3$. 
This ZPE should be compared with 
the energy gain from the JT coupling 
within the classical treatment of the O atoms. 
It is known from the quantum Monte Carlo calculation 
that the JT energy gain from the staggered and long-ranged distortion 
is the largest at half filling and becomes smaller with increasing 
hole-doping\cite{Motome_Imada_let,Motome_Imada_reg}. 
The JT energy gain and the ZPE 
become competing at about 10$\%$-doping, above which 
such long-ranged distortion melts. 
This is why it is necessary to realize 
short-ranged correlations of distortions in the calculations. 
Note here that the motion of the lattice has a time scale slower 
than that of the electrons.  
If one is on the viewpoint of the anti-adiabatic 
approximation\cite{anti_adiabatic}, 
interactions between the electrons and the lattice become 
instantaneous so that the effect of the lattice is renormalized 
into the Coulomb interaction, resulting in a simple complement 
to the physics of the Mott transition.  
On the other hand, the adiabatic treatment of the lattice 
can consider the difference between the two time scales 
as a limiting case of the retardation effect 
in the electron-phonon interaction. 
A treatment of the lattice fluctuation from this side 
of viewpoint would capture at least an aspect overlooked 
within the treatment of the Coulomb interaction only.  
In this paper, we treat the lattice classically and 
perform calculations simulating lattice fluctuation 
by taking Monte Carlo (MC) sampling 
at a given energy corresponding to the above ZPE. 

Now, we examine a model 
${\cal H}={\cal H}_{\rm el}+{\cal H}_{\rm JT}+{\cal H}_{\rm ph}$, 
where the first, second and third terms are the electronic part, 
the JT interaction between the lattice distortion and the 
electron, and the elastic term of the displacement 
of the oxygens, respectively.  
${\cal H}_{\rm el}$ 
is derived from the DE model 
with the $e_g$-orbital degeneracy 
under the strong Hund's-rule coupling and perfect spin polarization 
as $\sum_{ij}\sum_{\nu\nu^{\tiny \prime}}
t_{ij}^{\nu\nu^{\tiny \prime}} c^{\dagger}_{i\nu} 
c_{j\nu^{\tiny \prime}} + U \sum_{i}n_{i1}n_{i2}$ 
with the hopping integral $t_{ij}^{\nu\nu^{\tiny \prime}}$ 
and the effective interorbital Coulomb repulsion $U$. 
The orbitals $d_{x^2-y^2}$ and $d_{3 z^2-r^2}$ correspond to 
$\nu$\mbox{\,}$=$\mbox{\,}$1$ and 2, 
respectively. We consider the nearest-neighbor (NN) hopping  
$t_{ij}^{11} = -3/4 \tilde{t}_0$, 
$t_{ij}^{22} = -1/4 \tilde{t}_0$, 
$t_{ij}^{12} = t_{ij}^{21} = 
-(+)\sqrt{3}/4 \tilde{t}_0$ along
the $x$($y$)-direction.  
The hopping amplitude $\tilde{t}_0$ is modified 
by the displacement $u_{ij}$ of the O atom 
between the NN Mn atoms from the center   
according to the Harrison's law, that is 
$\tilde{t}_0 = t_0 (1-4u_{ij}^2/a^2)^{-7/2} $.  
The most important effect of three dimensionality is that 
the hopping in $z$ direction 
favors
OP into the $d_{3 z^2-r^2}$ orbital. 
This OP occurs because the transfer amplitude 
in the $z$ direction is larger for the $d_{3 z^2-r^2}$ orbital 
than $d_{x^2-y^2}$. 
To mimic this effect in 2D lattice, we add 
a lower chemical potential to $d_{3 z^2-r^2}$ orbital 
than $d_{x^2-y^2}$ so that a 3D isotropy of the OP 
is reproduced\cite{Nakano_Motome_Imada_orbital_U}. 
The added chemical potential can be viewed as the self-hopping term 
with an amplitude $\bar{t}_0$ for a single-layer system 
under the periodic boundary condition in $z$ direction. 
The JT interaction ${\cal H}_{\rm JT}$ is given by 
$-g \sum_{i} ( q_{3i}T_i^z+q_{2i}T_i^x ) $ 
where $\mbox{\boldmath $T$}_i$ is a pseudo-spin operator for orbitals. 
$q_{2i}$ ($q_{3i}$) is an amplitude corresponding to the $Q_2$ 
($Q_3$) normal mode in the distortions of the octahedra formed 
by the oxygens surrounding $i$-th Mn atom.  
The elastic term ${ H}_{\rm ph}$ is $k \sum_{i} \sum_{j}^{\prime}
u_{ij}^2$, with $j$ running over only NN sites of $i$-th site. 

The method employed in this work is the Lancz$\ddot{\rm o}$s 
exact diagonalization of the finite-size cluster 
of the above Hamiltonian.  
The system size is $\sqrt{10}\times\sqrt{10}$. 
The energy of the electronic part is minimized 
by choosing the boundary condition, where phase shift 
between two boundaries is optimized 
by introducing a flux. 
Note that such boundary condition 
successfully reduces the finite-size 
effect\cite{Nakano_Imada_usual-Hub}.  
In fact, one can see only small differences between results 
of $\sqrt{10}\times\sqrt{10}$ and $4\times 4$ 
in the previous study\cite{Nakano_Motome_Imada_orbital_U} 
without JT distortions.  
To realize the fluctuating lattice distortions, we 
combine the above exact diagonalization 
with the MC sampling. 
At each update, every oxygen is moved from its latest position 
to a place which  is uniformly random in a range 
with a certain width. It provides a possible new distortion pattern. 
Here, the width is adjusted as the acceptance ratio becomes about 70\%. 
Next, an exact diagonalization is performed 
for the new distortion pattern. 
The obtained total energy  $E_0^{\prime}$ is used to judge 
by the Metropolis algorithm of comparison with 
the Boltzmann weight $\exp (-E_0^{\prime}/E_0)$ 
determined from a given energy $E_0$ 
whether the new pattern is accepted or not. 
We interpret that $E_0$ characterizes the quantum zero-point energy. 
The updated pattern is used in the next MC step 
of the oxygen displacements.  
At each diagonalization step, $\sigma (\omega)$ is obtained from 
[$\sigma_{x} (\omega)$+$\sigma_{y} (\omega)$]/2, 
where  
$
\sigma_{\alpha} (\omega) = 2\pi e^2 D_{\alpha} \delta(\omega) 
+\frac{\pi e^2}{N}
{
\sum_{m(\ne 0)}}
\frac{|\langle m|j_{\alpha}|0\rangle |^{2}}{E^{\prime}_m-E^{\prime}_0} 
\delta(\omega -E^{\prime}_m+E^{\prime}_0)
$, whose incoherent part is calculated 
by the continued-fraction-expansion method\cite{frac_expansion}. 
Here, $D_{\alpha}$ is the Drude weight, $j_{\alpha}$ is a 
current operator along $\alpha$-direction 
($\alpha$={\it x},\mbox{\,}{\it y}) given by 
$
j_{\alpha} = - {\rm i}\sum 
t_{i,i+\delta_{\alpha}}^{\nu\nu^{\prime}}
(
 c^{\dagger}_{i\nu}                     c_{i+\delta_{\alpha} ,\nu^{\prime}}
- c^{\dagger}_{i+\delta_{\alpha} ,\nu^{\prime}} c_{i\nu}
) 
$, 
and $|m\rangle$ 
denotes an eigenstate of the system 
with the energy eigenvalue $E^{\prime}_m$. 
$|m$=0$\rangle$ represents the ground state. 
Note that the kinetic energy per site is given by 
$\frac{-4}{\pi {\rm e}^2}\int_0^{\infty}\sigma (\omega) {\rm d}\omega$. 
After repeating the 200 warm-up cycles starting 
from the mean-field solution, 
we calculate the MC average of $\sigma (\omega)$ 
by 1800 samplings. 
We have performed calculations at $g/t_0$=10 and $k/t_0$=100.  
These parameters are known to reproduce well the experimental results 
at half filling 
around $U/t_0$~$\sim$5\cite{Motome_Imada_let,Motome_Imada_reg}.
We here calculate $\sigma (\omega)$ for $U/t_0$=6 and 
the hole-doping concentration $\delta$ is 
0.2\cite{comment_delta0.2}, which are close values 
to the ones suggested from the experimental 
and previous theoretical works.  
In the above parameters, $\bar{t}_0$ is found to be $\sim$0.2  
when the mean-field solution is at $(q_{2i},q_{3i})\sim (0,0.031)$. 

In Fig. 1, the result of the distributions 
in the ($q_2,q_3$) plane is shown. 
One can see that 
when the given energy in the MC procedure 
is small ($E_0/t_0$=0.02), 
fluctuations of distortions are restricted 
only near the mean-field solution. 
On the other hand, 
when the given energy becomes larger ($E_0/t_0$=0.1), 
the distortions distribute in a larger area. 
This area includes (0,0) corresponding to the case where 
there are neither $Q_2$- nor $Q_3$-mode distortions. 
The system with the mean-field distortion has 
an equivalent point of 
a local energy minimum 
at a negative $q_{3i}$ with $q_{2i}$=0.  
In our simulation for $E_0/t_0$=0.1, 
distortions in the area between these points 
are densely produced, which results in 
the realization of a large fluctuation of the lattice. 

Incoherent parts of $\sigma(\omega)$ 
for two cases of the large and small fluctuations of the lattice 
together with the mean-field case are shown in Fig.~2. 
In each case of $E_0/t_0=0.1$, 0.02 and 0 (mean-field case), 
the weight has a minimum at about $\omega_{\rm c}/t_0 \sim 5$, above which 
the structure of the response to the upper Hubbard band 
due to the strong correlation appears. 
Incoherence below $\omega_{\rm c}$ corresponds 
to the lowest-temperature structure below about 1{\,}eV 
discussed in Ref.~\ref{Okimoto_reg}.  
Thus, $t_0$ is estimated as 0.2{\,}eV, which has an order consistent 
to the experimental indications\cite{Saitoh}.  
Note that with this estimation of $t_0$, $E_0$ for the case of 
Fig.~2(b) is $\sim$0.02 eV, which is in the range of 
the ZPE in the above argument. 
For $E_0/t_0=0.02$ in Fig.~2(a), 
the structure of $\sigma(\omega)$ is similar to that in the mean-field case 
with only small differences in the very low-energy region. 
The Drude weight is $D/t_0\sim0.036 \pm 0.001$ and 
the ratio $D/N_{\rm eff}\sim 0.35$, 
where $N_{\rm eff}$ is the effective carrier density 
defined as 
$\frac{1}{\pi {\rm e}^2}\int_0^{\omega_{\rm c}} \sigma(\omega) 
{\rm d}\omega$. 
The differences of these values from the mean-field case, 
$D/t_0\sim0.04$ and $D/N_{\rm eff}\sim 0.4$ are small. 
This indicates that small fluctuations 
do not fully reproduce the situation of the strong incoherence dominant 
in the charge dynamics as observed in the experiments. 
On the other hand, in the case 
of $E_0/t_0=0.1$ in Fig.~2(b), 
the Drude weight is $D/t_0 \sim 0.019 \pm 0.004$ and 
the ratio $D/N_{\rm eff}\sim 0.13$. 
This ratio is consistent with the experimental indication 
that the incoherence dominates the charge dynamics.  
Let us compare this Drude weight with the Drude weight  
for the free-electron system $D_{\rm FE} = \hbar^2 n/(2 m a^2)$, 
where $n$ is the electron density per unit cell. 
The ratio $D_{\rm FE}/D$ is estimated as 40 - 60, which agrees 
with 50 - 80 of the mass enhancement in the experiments. 
$\omega$-dependence of the incoherence below $\omega_{\rm c}$ shows 
a leaned shape on the low-energy region
and thus agrees with experimental indications. 
Increasing temperature changes this shape of the incoherent structure 
in experiments, 
which is a future problem because the present method is based on 
the Lancz$\ddot{\rm o}$s algorithm.  
Note in addition that the gap-like structure in the energy 
region lower than $\omega /t_0 \sim 1$ is seen in the mean-field case.   
A similar structure is considered to originate 
from the orbital-excitation gap due to the OP induced by 
the two dimensionality\cite{Horsch_gap}. 
Our calculations for the purely 2D case  with $\bar{t}_{0}=0$ 
show that the OP and the gap-like structure survive 
even when the lattice fluctuation is given from $E_{0}/t_{0}=0.1$. 
In the case of Fig.~2(b), on the other hand, 
the large lattice fluctuation 
forces the depolarized orbital, which results in the disappearance 
of the structure.  

This strong incoherence occurs as a result of 
the cooperative effect of the orbital degeneracy, 
the electron correlation and the largely fluctuating JT distortions. 
To see the importance of the complex fluctuations, 
let us compare this incoherence with cases of various amplitudes 
of the uniform $Q_3$ distortion 
including the mean-field solution (see Fig.~3).  
Among the cases of uniform $Q_3$ distortion,  
the charge dynamics is the most incoherent 
around the energy-optimization points for the mean-field solutions. 
In the larger amplitude of the uniform $Q_3$ distortion, 
the orbitals become polarized and 
either the orbital $d_{3z^2-r^2}$ or $d_{x^2-y^2}$ is favored. 
In the orbitally polarized state, electrons tend to move 
only within the favored orbital. 
This is why the charge dynamics recovers the coherence 
for the larger amplitude of uniform $Q_3$ distortion than the mean-field case. 
Therefore, it is understood 
that the strong incoherence obtained in the MC calculation 
for $E_0/t_0=0.1$ cannot be reached by 
only the uniform $Q_3$ distortion with any amplitude 
and 
that this strong incoherence is not realized 
only from a simple mixture of the cases 
with various amplitudes of uniform $Q_3$ distortions, 
one of which gives the mean-field solution with the largest 
JT energy gain. 

Finally, two controversial interpretations 
about the small Drude weight in experiments should be mentioned.  
Takenaka {\it et al}.\cite{Takenaka_x0.3,Takenaka_x0.175} made 
the measurement on cleaved surfaces and claimed that 
the coherence is not so small from the viewpoint 
of the obtained structure leaned on the low-energy region. 
From the measurement by Okuda {\it et al}.\cite{Okuda_NEC} 
after Takenakas' work, 
$\sigma(\omega)$ on the surface with annealing procedures 
removing the residual stress seems very similar 
to Takenakas' result. 
Note that all the experimental results show the common feature 
of this leaned shape up to $\sim$1\,eV, 
which is reproduced here in Fig.\,\mbox{2(b)}.  
However, Takenaka {\it et al}. did not estimate 
the true Drude response sharply centered at $\omega$=0 and 
instead assigned the broad incoherent response as the Drude weight,   
while the estimation of the Drude weight by Okuda {\it et al}. 
supports its smallness. 
Although Takenaka {\it et al}. attribute the controversy 
to the surface effect, our calculation suggests that 
pure two dimensionality recovers the coherence due to the increasing OP. 

In summary, we have studied the charge dynamics in the perovskite 
Mn oxides within the framework of 
the generalized double-exchange model including the three elements, 
namely, the orbital degeneracy, the electron correlation, and 
the Jahn-Teller coupling with largely fluctuating distortions.  
As the cooperative effects of all the elements, 
the optical conductivity, whose incoherence dominates 
the charge dynamics, has been obtained.  
The result is favorably compared with the incoherent charge dynamics 
observed in the experiments.  
Thereby it is found that 
not only the strong electron correlation in the orbital system 
but also the Jahn-Teller interactions with large fluctuations 
are indispensable 
in the minimal model describing the properties of the compounds. 
The manganite provides a suitable example where 
the interplay of all these elements is crucial 
not only in the undoped Mott insulator and the 
phase separated system in the very slightly doped 
case\cite{Motome_Imada_let,Motome_Imada_reg} 
but also in the doped metal near the Mott insulator.  
Contrary to the simple problem of random potential 
without electron correlations, electron motion has already been strongly 
obstructed by the correlation, 
under which the incoherence is triggered by the lattice fluctuation. 
Our results here give insights on such coupled effects under entanglement.
In this work, the lattice fluctuation is treated classically and 
simulated by making such situations in the Monte Carlo sampling.  
As a future problem, direct treatment of the quantum nature 
of the lattice is desired.

This work is supported by `Research for the Future Program' 
(JSPS-RFTF 97P01103) 
from the Japan Society for the Promotion of Science(JSPS). 
Y.M. is supported by Research Fellowships of the 
JSPS 
for Young Scientists.
A part of the computations was performed using the
facilities of the Supercomputer Center,
Institute for Solid State Physics, University of Tokyo.


\begin{thebibliography}{99}
\bibitem{comment_coefficient_phonon} 
Each oxygen atom is counted twice 
from the two neighboring Mn atoms; 
this is why a numerical coefficient in the momentum term is 1/4.
\bibitem{comment_delta0.2} 
In this doping concentration, the mixed boundary condition 
composed of the periodicity along one direction and 
the anti-periodicity along the other direction is chosen.
\end{thebibliography}
\end{document}